\documentstyle[twocolumn,prl,aps]{revtex}
\tighten
\begin{document}
%
%
\title{Onsager reciprocity relations without microscopic
reversibility}

\author{D. Gabrielli}
\address{SISSA, Scuola Internazionale Superiore di Studi Avanzati\\
Via Beirut 2-4, 34014 Trieste, Italia }
\author{G. Jona-Lasinio}
\address{Dipartimento di Fisica, Universit\`a di Roma "La Sapienza"\\
Piazza A. Moro 2, 00185 Roma, Italia}
\author{C. Landim}
\address{IMPA, Estrada Dona Castorina 110\\
J. Botanico, 22460 Rio de Janeiro RJ, Brasil\\
and LAMS de l'Universit\'e de Rouen, Facult\'e de Sciences\\
BP 118,F-76134 Mont-Saint-Aignan Cedex, France}

\maketitle

%
\begin{abstract}
In this paper we show that
Onsager--Machlup time reversal properties of thermodynamic fluctuations
and Onsager reciprocity relations for transport
coefficients can hold also if the
microscopic dynamics is not reversible.
This result is based on
the explicit construction of a class of conservative models which can
be analysed rigorously.
\end{abstract}
%
%
%
%


\def\N{I\!\! N}
\def\Z{Z\!\!\! Z}
\def\R{I\!\! R}
\def\T{{T\!\!\! T}}


\narrowtext

\medskip

Fundamental contributions to the theory of irreversible processes
were the derivation of the reciprocal relations for transport
coefficients in states deviating only slightly from equilibrium
and the calculation of the most probable trajectory creating
a fluctuation near equilibrium. The first result was obtained
by Onsager in 1931 \cite{ON1} and the second one by Onsager
and Machlup \cite{ON2} in 1953. The calculation of the most
probable trajectory relies on the reciprocal relations which
in turn are a consequence of microscopic reversibility. It turns
out that the trajectory in question is just the time reversal
of the most probable trajectory describing relaxation to equilibrium
of a fluctuation. The latter is a solution to the hydrodynamical
equations.

These topics have received a certain amount of attention in
the physical litterature in the course of the last forty years.
No rigorous results have been however established.
More recently, this subjet has been taken
up in various papers attempting more rigorous approaches~:
\cite{E}, \cite{GJLV}, \cite{ELS}, in the context
of so called interacting particle systems and \cite{G1},
\cite{G2} in a context of deterministic dynamical systems.

In \cite{GJLV} we discussed the following question: is  microscopic
reversibility a necessary condition for the validity of the
Onsager and Onsager--Machlup results? The answer to this
question is far from obvious
because in going from the microscopic to the macroscopic scale
a lot of information is lost and irreversibilities at a small scale
may be erased when taking macroscopic averages.

In \cite{GJLV} we have exhibited a class of
microscopic nonreversible stochastic dynamics for which
the time reversal rule of Onsager--Machlup is still valid
even for fluctuations very far from equilibrium.
This class of dynamics concerns dissipative one component systems with
a hydrodynamic equation of
gradient type. Therefore there is no Onsager reciprocity relation
to verify.

In this paper we present a class of nonreversible multi component
conservative models giving rise at
the macroscopic level to nonlinear purely diffusive equations
in the terminology of \cite{E}.

The equations are of the following form
\begin{equation}
\partial_t \rho = \sum_{i=1}^d \partial_{u_i}
\Big\{ D(\rho) \cdot \partial_{u_i} \rho\Big\}
\end{equation}
where $\rho(u,t) =(\rho_1(u,t),\dots , \rho_n(u,t))$ is
a vector standing for the densities of different kinds
of particles and $D$ is in general a nonsymmetric $n\times n$ matrix.

Associated to our models there is an entropy functional
$S(\rho)$ that is written as the integral of a
density $s(\rho)$~: $S(\rho) = \int s(\rho(u)) du$.

The Onsager coefficients are defined in this context by
\begin{equation}
L (\rho) = D(\rho) \cdot R(\rho)
\end{equation}
where the matrix $R$ is determined by the entropy density
$s(\rho(u))$ in the following way
\begin{equation}
(R^{-1})_{i,j} = \frac{ \partial^2}{\partial \rho_i(u) \partial \rho_j(u)}
s(\rho(u))
\label{rr}
\end{equation}
which is by definition a symmetric matrix.
Onsager's reciprocity relations mean that
$L$ is a symmetric matrix, a property which holds
for our models.

In the physical literature one usually proves the
Onsager--Machlup time reversal property from Onsager
reciprocity relations. In our approach, we follow the opposite
order~: we obtain a large deviation functional from which
we prove the Onsager--Machlup time reversal property and compute
the entropy.
This in turn allows us to prove Onsager reciprocity
relations.
Our results are not restricted
to the neighborhood of the equilibrium.

For simplicity, we shall restrict ourselves to one dimensional
two component models but all analysis can be carried out
for any space dimension or for any number of components.
As in \cite{GJLV}, we consider periodic boundary conditions.
The systems considered in this paper differ from those
of \cite{GJLV} due to the conservative character of the
dynamics.

We consider an interacting particle system that
describes the evolution of two types of particles on a lattice.
The stochastic dynamics can be informally described
as follows. Fix a nonnegative function
$g: \N\rightarrow \R_+$ such that $g(0)=0<g(i)$
for $i\ge 1$ and a finite range mean zero
transition probability $p(\cdot)$ on $Z\!\!\! Z$ ($\sum_y
y p(y) =0$ and $p(x)=0$ for $|x|$ large enough).
We shall assume the jump rate $g$ to be Lipschitz and
to diverge at the boundary~: $|g(k+1)-g(k)| \le l_0$
and $\lim_{k\to\infty} g(k) =\infty$.
If there are $k_i$, $i=1,2$, particles of type $i$ at a site
$x$ of $\Z$, at rate $p(y) g(k_1+k_2) \{k_i/k_1+k_2\}$
one particle of type $i$ jumps from site $x$ to $x+y$.
This happens independently at each site.

To define precisely the process, we introduce some
notation. We shall consider particles evolving on the
discrete one dimensional torus with $N$ points, denoted
by $\Z_N$ (the integers modulo $N$). Sites of $\Z_N$ are denoted
by $x$, $y$ and the configurations by the greek letter $\eta=(\eta_1,\eta_2)$
so that $\eta_i (x)$ stands for the total number of particles
of type $i$ at site $x$ for the configuration $\eta$.

The generator $L_N$ of this Markov process acts on
functions $f$ as
\begin{equation}
L_Nf =\frac{N^2}{2}\sum_{j=1}^2 \sum_{x,y\in \Z_N}
p(y) T_j^{x,x+y} f
\label{eqn1}
\end{equation}
where the addition in $\Z_N$ means addition modulo $N$;
the operators $T_1^{x,y}$ are defined by
$$
(T_1^{x,y} f) (\eta_1,\eta_2) = r_{x}(\eta)
\eta_j (x) [f(\eta_1^{x,y}, \eta_2) - f(\eta_1, \eta_2)]
$$
with $r_{x}(\eta) = g(\eta_1(x)+\eta_2 (x))
/\{\eta_1(x)+\eta_2 (x)\}$ and
$\zeta^{x,y}$ is the configuration obtained from $\zeta$
letting one particle jump from $x$ to $y$~:
\begin{equation}
\zeta^{x,y}(z)= \left\{
\begin{array}{ccl}
\zeta(z) &\hbox{if}& z\neq x,y \\
\zeta(z) -1 &\hbox{if}& z=x \\
\zeta(z) +1  &\hbox{if}& z=y \; .
\end{array}
\right.
\label{eqn2}
\end{equation}
The operators $T_2^{x,y}$ are defined in a similar way.

This process has two conserved quantities~: the total
number of $\eta_1$-particles and the total number of
$\eta_2$-particles. It is therefore expected that for
each fixed density $\rho_i\ge 0$ there
should exist an equlibrium state with global
density of $\eta_i$-particles equal to $\rho_i$.

To describe these equilibrium probability measures,
for each $\varphi_1,\varphi_2 \ge 0$, consider the product
probability measure $\nu^N_{\varphi_1,\varphi_2}$ on
$\N^{\Z_N}\times \N^{\Z_N}$ defined by
\begin{equation}
\begin {array}{ccl}
& \nu^N_{\varphi_1,\varphi_2} \{(\eta,\xi);\; \eta (x)=k_1, \xi (x)=k_2\} \\
&\quad = \frac{1}{Z(\varphi_1,\varphi_2)}
\frac{\varphi_1^{k_1} \varphi_2^{k_2}}
{g(k_1+k_2)!} \frac{ (k_1+k_2)!}{k_1! k_2!}
\end{array}
\label{eqm}
\end{equation}
for $k_1\ge 0$ and $k_2\ge 0$. In this formula
$Z(\varphi_1,\varphi_2)$ is a normalizing constant and
$g(k)!$ stands for $g(1)\cdots g(k)$. A simple computation shows that
these measures are invariant for the Markov process with generator
$L_N$ and reversible if and only if the transition probability
$p(\cdot)$ is an even function.

Define $\rho_i: \R_+\times\R_+\to\R_+$ by $\rho_i(\varphi_1,\varphi_2) =
E_{\nu^N_{\varphi_1,\varphi_2}} [\eta_i (0)]$ and set
$\rho^* = \rho_1 +\rho_2$. One can check that $\rho^*$ is a
smooth strictly increasing function of $\varphi_1 +\varphi_2$.
Denote  by $a = a(\rho^*)$ the inverse of $\rho^*$~:
$a(\cdot) = (\rho^*)^{-1} (\cdot)$. We have that
$\varphi_i = (\rho_i/\rho_1+\rho_2) a(\rho_1+\rho_2)$.
To keep notation simple we shall denote by $b (\rho^*)$ the
function $a(\rho^*)/\rho^*$. In conclusion, for each fixed
density $(\rho_1,\rho_2)$ we obtained an invariant state
with total density of $\eta_i$-partciles equal to $\rho_i$.
We shall from now on fix  a density $\bar\rho =
(\bar\rho_1,\bar\rho_2)$.

Let us consider now the unit interval $\T =[0,1)$ with periodic boundary
condition and functions $\gamma_i: \T\to \R_+$, $i=1,2$ with
global density $\bar\rho_i$~: $\int_{\T} \gamma_i (u) du =\bar\rho_i$.
The main object of our study is the empirical density $\mu^N (t)
=(\mu^N_1 (t), \mu_2^N (t))$~:
\begin{equation}
\mu^N_i(t,x)=\frac{1}{N}\sum_{y\in \Z_N}\eta_i(t,y)\delta(x-\frac{y}{N}) .
\label{eqn5}
\end{equation}
If we denote by $Q_{\gamma_1,\gamma_2}^N$ the distribution law
of the tragectories $\mu^N(t)$  when the initial measure
is concentrated on a configuration pair $(\eta^N_1,\eta_2^N)$
such that $\mu^N(0)\rightarrow (\gamma_1(u) du, \gamma_2(u) du)$
as $N\uparrow \infty$,
it is possible to show that
$Q_{\gamma_1,\gamma_2}^N$ converges weakly as $N$ goes to infinity
to the  measure concentrated on
the path $\rho(u,t)$ that is the unique solution of
\begin{equation}
\left\{
\begin{array}{ll}
\partial_t \rho = (\sigma^2/2) \partial_u
\Big\{ D (\rho) \cdot \partial_u \rho \Big\} & \\
\rho(0,\cdot) = \gamma(\cdot)\; . &
\end{array}
\right.
\label{eqn6}
\end{equation}
In this formula $\sigma^2 = \sum_y y^2 p(y)$ and
$D=D(\rho)$ is the nonsymmetric diffusion matrix
given by
\begin{equation}
D(\rho) = b (\rho) I +b'(\rho) J(\rho)\; ,
\label{difmat}
\end{equation}
where $I$ is the identity and $J(\rho)$ is the matrix
with entries $J_{i,j}(\rho) =\rho_i$.

The above result is a law of large numbers that shows that the empirical
density in the limit of large $N$ behaves deterministically
according to equation (\ref{eqn6}).
We can now ask what is the probability that our system follows a trajectory
different from the solution of (\ref{eqn6}) when $N$ is large
but not infinite. This probability is exponentially small in $N$ and
can be estimated using the methods of the theory of large deviations
introduced for the systems of interest in \cite{KOV} and \cite{DV}.
The main idea consists in introducing a modified
system for which the trajectory of interest (fluctuation) is typical
being a solution of the corresponding hydrodynamic equation, and then
comparing the two evolutions. For this purpose, for each
pair of smooth functions $H_i=H_i(u,t)$, we consider the
Markov process defined by the generator
$$
L_{N,t}^H f =\frac{N^2}{2}\sum_{j=1}^2
e^{H_j((y+x)/N, t)-H_j(x/N,t)} p(y) T_j^{x,x+y} f
$$
with $p(\cdot)$ and $T_j^{x,x+y}$ as previously defined.
The function $H$ can be interpreted as an external field.

The deterministic equation satisfied by the empirical density is now
\begin{equation}
\left\{
\begin{array}{ll}
\partial_t \rho = (\sigma^2/2)
\partial_u \Big\{ D (\rho) \cdot \partial_u \rho \Big\} -
\sigma^2 \partial_u \Big\{ b (\rho) A(\rho, H) \Big\}& \\
\rho(0,\cdot) = \gamma(\cdot) &
\end{array}
\right.
\label{eqn10}
\end{equation}
where $A(\rho,H)$ is the vector with components $A_i = \rho_i \partial_u H_i$.

Given a function $\rho(u,t)$ twice differentiable with respect to $u$ and once
with respect to $t$ and such that $\int_\T \rho_i(u,t) du = \bar\rho_i$
this equation determines uniquely the field $H=(H_1,H_2)$.

The probability that the original system
follows a trajectory different from a solution of (\ref{eqn6})
can now be expressed in terms of the
field $H$. We introduce the large deviation functional
\begin{equation}
I_{0,t_0} (\rho) = \sum_{i=1}^2 (\sigma^2 /2)
\int_0^{t_0}dt \, \int_\T du\,
b (\rho) \rho_i (\partial_u H_i)^2\; .
\label{eqn14}
\end{equation}

Let $\cal G$ be a set of trajectories in the interval $[0,t_0]$.
The large fluctuation estimate asserts that
\begin{equation}
Q_{\gamma_1,\gamma_2}^N({\cal G})\simeq e^{-NI_{0,t_0}({\cal G})}
\label{lf}
\end{equation}
where
\begin{equation}
I_{0,t_0}({\cal G})=\inf_{\rho \in {\cal G}}I_{0,t_0}(\rho)
\label{inf}
\end{equation}
The sign $\simeq$ has to be interpreted as asymptotic equality
of the logarithms.

{}From the equations (\ref{lf}), (\ref{inf}), one sees that to find
the most probable trajectory that connects the equilibrium
$\bar \rho$ to a certain state $\gamma(u)$
one has to find the $\rho(u,t)$ that minimizes
$I_{-\infty, 0}(\rho)$ in the set
${\cal G}_\gamma$ of all trajectories satisfying the boundary
conditions
\begin{equation}
\lim_{t\rightarrow -\infty}\rho(u,t)=\bar \rho\; ,\quad
\rho(u,0)=\gamma(u)\; .
\label{eqn32}
\end{equation}

It is now possible
to prove, following the same approach of \cite{GJLV},
that the unique solution of
our variational problem is the function $\rho^*(u,t)$ defined by
\begin{equation}
\rho^*(u,t)=\rho(u,-t)
\label{eqn35}
\end{equation}
where $\rho(u,t)$ is the solution of the hydrodynamic
equation which relaxes
to equilibrium with initial state $\gamma$. $\rho^*(u,t)$
is therefore a solution of the hydrodynamic
equation with inverted drift
\begin{equation}
\partial_t\rho=-(\sigma^2/2)  \partial_u
\Big\{ D (\rho) \cdot \partial_u \rho \Big\}\; .
\label{eqn36}
\end{equation}
Equation (\ref{eqn35}) is the Onsager-Machlup time-reversal relation.

Denote by $S(\gamma)$ the functional defined by
\begin{equation}
S(\gamma)=\inf_{\rho\in{\cal G}_\gamma} I_{-\infty, 0}(\rho)\; .
\label{entropy}
\end{equation}
which, by the Boltzmann--Einstein relationship,
has to be identified with the entropy of the system.
By inserting (\ref{eqn35}) in (\ref{entropy}) we obtain
an explicit formula for the entropy~:
$$
S(\gamma) = \int_\T s(\gamma (u)) du
$$
where
$$
\quad s(\gamma) = \sum_{j=1}^2 E(\gamma_j (u))  +
F(\gamma_1(u) + \gamma_2(u) )
$$
and $E(\rho) = \int^\rho \log \rho' d\rho'$,
$F(\rho) = \int^\rho \log b(\rho') d\rho'$.
It is possible to check that $S(\rho(\cdot ,t))$ decreases in time if
$\rho(\cdot,t)$ is a solution of the hydrodynamic
equation (\ref{eqn6}).

Of course, the entropy could also be calculated
from the equilibrium measure (\ref{eqm}) and it is easy to
see that the two expressions coincide up to an additive constant.

This explicit expression for the entropy
$S(\cdot)$ permits to check Onsager's relations
in our model. A simple computation shows that
the matrix $R$ defined by equation (\ref{rr}) is
such that
\begin{equation}
(R^{-1})_{i,j} = \delta_{i,j} \frac{1}{\gamma_i(u)} +
\frac{b' (\gamma (u))}{b(\gamma (u))}
\end{equation}
where $\delta_{i,j}$ stands for the delta of
Kronecker and $s(\gamma)$ is the entropy density.
The product $L=DR$ can now be computed using the
explicit formula for $D$ given in
(\ref{difmat}) and shown to
be a symmetric matrix.

In all the above calculations we never used
the symmetry properties of the transition probability
$p(\cdot)$ so that they are valid both for reversible
and irreversible dynamics.

This provides conclusive evidence that macroscopic reversibility,
in the sense of validity of the above results,
does not require microscopic reversibility.

\acknowledgments
G. J-L. acknowledges a useful correspondence with G. Eyink
and interesting discussions with G. Gallavotti.
C. L., thanks the INFN (sezione di Roma)
and the CNRS-CNR agreement for hospitality and support.

%
%

%
%

\end{document}